\documentstyle[11pt]{article}
\begin{document}
\title{Wigner Functions in Curved Space--Time
and Quantum Corrections to Thermal Equilibrium}
\author{Oleg A. Fonarev\thanks{talk given at the 3rd International Wigner
Symposium, 5th-11th September 1993, Oxford, UK.}
\\ {\em Racah Institute of Physics, The Hebrew University}
\\ {\em Jerusalem 91904, Israel}
\\ {\em E--mail: OLG@vms.huji.ac.il}}
\date{}
\maketitle

\hspace*{.3in}
The Wigner function is known to be a very useful tool in exploring the
semiclassical limit of the quantum theory. It is defined as a Fourier
representation of the density matrix ${\bf \rho}$ with respect to the
difference of coordinates$^{\ref{kn:wigner}}$:
\begin{equation}
{f}_{w}(q,p)=(\pi\hbar)^{-1}\:\int_{- \infty}^{\infty}
dQ\:e^{-2iQp/\hbar}\langle q+Q |{\bf \rho}|q-Q\rangle \; . \label{eq:wigo}
\end{equation}
At a high temperature or after coarse graining, the Wigner function gives us
directly quantum corrections to a distribution function in the classical phase
space (the Boltzmann distribution if the system is in thermal equilibrium). \\
\hspace*{.3in}
In curved space--time the notion of the Fourier transformation becomes
ambiguous because it is not invariant under general coordinate transformations.
Different approaches to solve this problem have been
proposed$^{\ref{kn:winter}-\ref{kn:fonarev1}}$. I will follow the approach by
[\ref{kn:fonarev1},\ref{kn:fonarev11}] which uses the formalism of the tangent
bundles. Consider any point $x$ on a space--time manifold ${\cal M}$. Let
${{\bf \varphi}}(x)$ be a (real) scalar field on the manifold. Let us introduce
the tangent space ${\cal T}_{x}(\cal M)$ at point $x$ and define field ${{\bf
\Phi}}(x,y)$ on the tangent bundle as follows:
\begin{equation}
{{\bf \Phi}}(x,y) = \exp(y^{\alpha} \hat{\nabla}_{\alpha}) \, {{\bf
\varphi}}(x) \; .       \label{eq:Phi}
\end{equation}
Here the operator
$\hat{\nabla}_{\alpha} = \nabla_{\alpha} - \Gamma_{\alpha\gamma}^{\beta} \:
y^{\gamma} \: \frac{\partial}{\partial y^{\beta}}$
is the horizontal lift of the derivative operator to the tangent
bundle$^{\ref{kn:yano}}$. The generalized Wigner function is now defined as
follows:
\begin{equation}
{f}(x,p) = (\pi\hbar)^{-4}\: \sqrt{-{g}(x)}\:\int_{{\cal T}_{x}(\cal M)}
d^{4}\,y\:e^{-2iy^{\alpha}p_{\alpha}/\hbar}\:\langle {{\bf \Phi}}(x,-y){{\bf
\Phi}^{\dagger}}(x,y)\rangle \; . \label{eq:wigf}
\end{equation}
Here the brackets mean the averaging with a density matrix on an initial Cauchy
hypersurface. In the Minkowski space--time, the tangent space ${\cal
T}_{x}(\cal M)$ coincides with the base space ${\cal M}$, the exponent in
Eq.(\ref{eq:Phi}) becomes the shift operator, i.e. ${{\bf \Phi}}(x,y) = {{\bf
\varphi}}(x+y)$ and we get the usuall definition of the relativistic Wigner
function$^{\ref{kn:degroot}}$.
In curved space--time, the definition (\ref{eq:wigf}) is explicitly covariant
and allows us to easily derive equations describing the evolution of the
generalized Wigner function in the phase space, given a field equation for the
field ${{\bf \varphi}}$. \\
Suppose that ${{\bf \varphi}}$ obeys the covariant Klein--Gordon equation with
the conformal coupling to gravity:
\begin{equation}
\left(\nabla^{\alpha} \nabla_{\alpha} + m^{2}/\hbar^{2} -
\frac{1}{6} {R}\right){{\bf \varphi}} = 0 \; ,          \label{eq:Klein}
\end{equation}
where $R$ is the scalar curvature. We then get two equations for the Wigner
function$^{\ref{kn:winter}-\ref{kn:fonarev11}}$:
\begin{eqnarray}
\left( m^{2} - p^{\alpha}p_{\alpha} \right) \:
f =
- \frac{\hbar^{2}}{4}\, D^{\alpha} D_{\alpha} \: f - \hspace*{1in} \nonumber \\
\mbox{} - \hbar^{2} \left( \frac{1}{6} R + \frac{1}{12} R_{\alpha\mu\beta\nu}
p^{\mu} p^{\nu} \frac{\partial^{2}}{\partial p_{\alpha} \partial p_{\beta}} +
\frac{1}{4} R_{\mu\nu} p^{\mu} \frac{\partial}{\partial p_{\nu}} \right) \: f +
{O}(\hbar^{4}) \; , \label{eq:mashf}
\end{eqnarray}
\begin{eqnarray}
p^{\alpha} D_{\alpha} \, f =
\hbar^{2} \left(
 \frac{1}{6} R_{\nu\beta\mu\alpha} p^{\mu} \frac{\partial^{2}}{\partial
p_{\alpha} \partial p_{\beta}} D^{\nu} -\frac{1}{24}
R_{\alpha\mu\beta\nu ; \sigma} p^{\mu} p^{\nu} \frac{\partial^{3}}{\partial
p_{\alpha} \partial p_{\beta} \partial p_{\sigma}} +  \right. \nonumber \\
\left.
\mbox{} + \frac{1}{12} R_{\alpha}^{\nu} \frac{\partial}{\partial p_{\alpha}}
D_{\nu}
 - \frac{1}{24} R_{\alpha\beta ; \nu} p^{\nu} \frac{\partial^{2}}{\partial
p_{\alpha} \partial p_{\beta}} - \frac{1}{24} R_{; \alpha}
\frac{\partial}{\partial p_{\alpha}} \right)\: f + {O}(\hbar^{4}) \; ,
\hspace*{.1in} \label{eq:tranf}
\end{eqnarray}
where $
D_{\alpha} =  \nabla_{\alpha}
+ \Gamma_{\alpha\beta}^{\gamma}\: p_{\gamma}\: \frac{\partial}{\partial
p_{\beta}}$
is the horizontal lift of the derivative operator to the {\em cotangent}
bundle$^{\ref{kn:yano}}$. \\
Eq.(\ref{eq:mashf}) is a generalized mass--shell constraint, and it implies
that in the semiclassical limit, the structure of the Wigner function must be
as follows$^{\ref{kn:hu}}$:
\begin{equation}
f = F_{0} \, {\delta}(m^{2}-p^{2}) +
\hbar^{2} F_{1} \, {\delta'}(m^{2}-p^{2}) +
\hbar^{2} F_{2} \, {\delta''}(m^{2}-p^{2}) + {O}(\hbar^{4}) \; .
\label{eq:expf}
\end{equation}
All the $F_{n}$'s in (\ref{eq:expf}) are easily found from eq.(\ref{eq:mashf})
and they are expressed$^{\ref{kn:hu},\ref{kn:fonarev11}}$ in terms of certain
local differential operators acting to the function $F_{0}$. The function
$F_{0}$ satisfies the quantum corrected Vlasov equation (\ref{eq:tranf}) on the
mass--shell. $F_{0}$ is decomposed to classical and quantum parts:
\begin{equation}
F_{0} = F_{cl} + \hbar^{2} F_{qu} + {O}(\hbar^{4}) \; ,   \label{eq:F0}
\end{equation}
where $F_{qu}$ is found by integrating the classical distribution function
$F_{cl}$ along classical trajectories and is in general nonlocal in the phase
space. \\
Thus, given a classical distribution function ${F_{cl}}(x,p)$ as a solution to
the Vlasov equation, one is able, in principle, to evaluate the semiclassical
expansion of the generalized Wigner function and of any physical observable
expressible in terms of the Wigner function. For example, the number--flux
${N_{\alpha}}(x)$ and the stress--energy vector ${T_{\alpha\beta}}(x)$ are
given by$^{\ref{kn:winter},\ref{kn:fonarev11}}$:
\begin{equation}
{N_{\alpha}}(x) = \int\frac{d^{4}\,p}{\sqrt{-{g}(x)}}\: p_{\alpha}\:f(x,p) \; ,
       \label{eq:flux}
\end{equation}
\begin{eqnarray}
{T_{\alpha\beta}}(x) =
\int\frac{d^{4}\,p}{\sqrt{-{g}(x)}}\: p_{\alpha}\,p_{\beta}\:{f}(x,p) +
\hspace*{1in} \nonumber \\
 \mbox{} + \hbar^{2}\left(\frac{1}{6} {R_{\alpha\beta}}(x) + \frac{1}{12}
(\nabla_{\alpha}\nabla_{\beta} -
{g_{\alpha\beta}}(x)\nabla^{\nu}\nabla_{\nu})\right)
\int\frac{d^{4}\,p}{\sqrt{-{g}(x)}}\: {f}(x,p)
\; . \label{eq:ener}
\end{eqnarray}
\hspace*{.3in} Let us now consider a stationary space--time admitting a global
time--like Killing vector $\xi^{\alpha}$. In that case, any function,
${F_{cl}}(E)$, of the constant of motion, $E=\xi^{\alpha} p_{\alpha}$, is a
solution of the Vlasov equation. The classical stress--energy tensor calculated
with such distribution has a perfect fluid's structure.
The quantum corrections of the lowest adiabatic order to this distribution were
found in Ref.[\ref{kn:fonarev2}] and they possess the following properties:
a) all the quantum corrections are local in the phase space and they are
obtained by acting a certain differential operator to $F_{cl}$. This means that
physical observables are expressed in terms of momenta of the classical
distribution function;
 b) the stress--energy tensor has the following structure:
\begin{equation}
T_{\alpha\beta} = (\varepsilon + p) v_{\alpha} v_{\beta} - p g_{\alpha\beta} -
\Pi_{\alpha\beta} \; , \label{eq:quen}
\end{equation}
which is not longer a perfect fluid's one : the anisotropic pressure tensor
$\Pi_{\alpha\beta}$ appears ($\Pi^{\alpha}_{\alpha} = v^{\alpha}
\Pi_{\alpha\beta} = 0$);
c) the eigenvector of the stress--energy tensor, $v_{\alpha}$, doesn't coincide
with the hydrodynamical velocity $u_{\alpha} = N_{\alpha}
(N^{\nu}N_{\nu})^{-1/2}$. This leads to the nonvanishing heat flow:
$I_{\alpha} = (\varepsilon + p) (v_{\alpha} - u_{\alpha})$.
$I_{\alpha}$ is proportional to the combination $N_{0}/6N_{2} - N_{1}/4N_{3}$,
where $N_{k} = \int_{0}^{\infty} dE E^{k} {F_{cl}}(E)$. For the Boltzmann or
Bose--Einstein distributions this combination doesn't vanish. \\
Only in static space--times (i.e. for nonrotating systems) the quantum
corrections preserve the perfect fluid's structure with the vanishing heat
flow.
\\
\hspace*{.3in} Let us next consider a more general case of conformally
stationary (or conformally static) space--times. The definition (\ref{eq:wigf})
of the generalized Wigner function given allows one to easily link Wigner
functions in two conformally related manifolds. Under the conformal
transformation, ${g_{\alpha\beta}}(x) \mapsto {a}(x)^{2} {g_{\alpha\beta}}(x),
{\varphi}(x) \mapsto {\varphi}(x)/{a}(x)$, the Wigner function transforms as
follows$^{\ref{kn:fonarev3}}$:
\begin{equation}
f \mapsto a^{2} (1 + \hbar^{2} \hat{A} + {O}(\hbar^4)) f \; ,
\end{equation}
where $\hat{A}$ is a certain differential operator involving derivatives of the
conformal factor ${a}(x)$. If one knows a Wigner function in a stationary
space--time, one is able to compute a Wigner function in a conformally
stationary space--time. We have found$^{\ref{kn:fonarev3}}$ by this method the
quantum corrections to a class of distribution functions isotropic in the
momentum space$^{\ref{kn:ehlers}}$, $F_{cl}={F}(u^{\alpha}p_{\alpha},x)$.
\\
\hspace*{.3in} The Robertson--Walker Universes are examples of conformally
static (and also conformally flat) space--times. The metric tensor takes the
form: $g_{\alpha\beta}={a}(t)^2 \gamma_{\alpha\beta}$ where
$\gamma_{\alpha\beta}$ is the metric tensor of a spherically symmetrical static
space--time, and $t$ is the conformal time. The classical distribution function
compatible with the symmetry of the space--time is any function of the
three--momentum$^{\ref{kn:ehlers}}$.
The quantum corrections to such distributions were first found by
[\ref{kn:pirk}] through a different method. The structure of the
quantum--corrected stress--energy tensor is a perfect fluid's one in this case.
We also found$^{\ref{kn:fonarev3}}$ that the quantum corrected energy density
takes the following form:
\begin{equation}
{\varepsilon}(t) = {\varepsilon_{cl}}(t) + \frac{1}{\kappa}
\frac{m^{2}}{M_{pl}^{2}} {\cal K}(\frac{m}{{\Theta}(t)}) {G_{0}^{0}}(t) +
{O}(\hbar^{4})  \label{eq:enRW}
\end{equation}
where ${\cal K}$ is a certain function of the ratio of the particles mass $m$
to the local temperature ${\Theta}(t)=T/{a}(t)$, $M_{pl}$ is the Planck mass
and $G_{0}^{0}$ is the zero--zero component of the Einstein tensor. If one
substitutes Expr.\ (\ref{eq:enRW}) into the Einstein equation, one finds that
the effective gravitational "constant" appears which varies with time:
\begin{equation}
\frac{1}{{\kappa_{eff}}(t)} = \frac{1}{\kappa} (1 - \frac{m^{2}}{M_{pl}^{2}}
{\cal K}(\frac{m}{{\Theta}(t)})) \; .  \label{effgr}
\end{equation}
The variation is negligible at present but it could lead to important physical
effects in the early Universe.

\vspace{1\baselineskip}
\newcounter{refer}
\begin{list}{$^{\arabic{refer}}$}{\usecounter{refer}
\setlength{\rightmargin}{\leftmargin} \footnotesize}
\item \label{kn:wigner} E. P. Wigner, Phys. Rev. {\bf 40}, 749 (1932).
\item \label{kn:winter} J. Winter, Phys. Rev. D {\bf 32}, 1871 (1985).
\item \label{kn:hu} E. Calzetta, S. Habib, and B. L. Hu, Phys. Rev. D {\bf 37},
2901 (1988).
\item \label{kn:kandrup} H. E. Kandrup, Phys. Rev. D {\bf 37}, 2165 (1988).
\item \label{kn:fonarev1} O. A. Fonarev, Izv. VUZov. Fizika (USSR) {\bf 9}, 47
(1990).
\item \label{kn:fonarev11} O. A. Fonarev, preprint gr--qc/9309005; to appear in
J.Math.Phys. (USA).
\item \label{kn:yano} K. Yano and Sh. Ishihara, {\em Tangent and Cotangent
Bundles: Differential Geometry} (Pure and Applied Mathematics: A Series of
Monographs and Textbooks) (Marcel Dekker, Inc.\ , New York, 1973).
\item \label{kn:degroot} S. R. de Groot, W. A. van Leeuwen, and Ch. G. van
Weert, {\em Relativistic Kinetic Theory} (North--Holland, Amsterdam, 1980).
\item \label{kn:fonarev2} O. A. Fonarev, Phys. Lett. A {\bf 152}, 153 (1991).
\item \label{kn:fonarev3} O. A. Fonarev, in preparation.
\item \label{kn:ehlers} J. Ehlers, P. Geren and R. K. Sachs, J. Math. Phys.
{\bf 9}, 1344 (1968).
\item \label{kn:pirk} K. Pirk and G. B$\ddot{o}$rner, Class. Quantum Grav. {\bf
6}, 1855 (1989).
\end{list}
\end{document}